\documentclass[9pt,conference]{IEEEtran}
\IEEEoverridecommandlockouts
\usepackage{cite}
\usepackage{amsmath,amssymb,amsfonts}
\usepackage{graphicx} 
\usepackage{algorithmic}
\usepackage{graphicx}
\usepackage{textcomp}
\usepackage{xcolor}
\usepackage{float}
\usepackage{algorithm}
\usepackage{setspace}
\let\Algorithm\algorithm

\renewcommand\algorithm[1][]{\Algorithm[#1]\setstretch{1.2}}
\usepackage{multirow}
\usepackage{subcaption}
\usepackage{mwe}
\def\BibTeX{{\rm B\kern-.05em{\sc i\kern-.025em b}\kern-.08em
    T\kern-.1667em\lower.7ex\hbox{E}\kern-.125emX}}

\newtheorem{theorem}{Theorem}

\begin{document}

\title{Weighted Multi-projection: 3D Point Cloud Denoising with Estimated Tangent Planes \\
}

\author{\IEEEauthorblockN{Chaojing Duan}
\IEEEauthorblockA{\textit{Electrical \& Computer Engineering} \\
\textit{Carnegie Mellon University}\\
Pittsburgh, USA \\
chaojind@andrew.cmu.edu}
\and
\IEEEauthorblockN{Siheng Chen}
\IEEEauthorblockA{\textit{Uber ATG} \\
Pittsburgh, USA \\
sihengc@uber.com}
\and
\IEEEauthorblockN{Jelena Kova\v cevi\' c}
\IEEEauthorblockA{\textit{Electrical \& Computer Engineering} \\
\textit{Carnegie Mellon University}\\
Pittsburgh, USA \\
jelenak@cmu.edu}
}

\maketitle

\begin{abstract}
As a collection of 3D points sampled from surfaces of objects, a 3D point cloud is widely used in robotics, autonomous driving and augmented reality. Due to the physical limitations of 3D sensing devices, 3D point clouds are usually noisy, which influences subsequent computations, such as surface reconstruction, recognition and many others. To denoise a 3D point cloud, we present a novel algorithm, called weighted multi-projection. Compared to many previous works on denoising, instead of directly smoothing the coordinates of 3D points, we use a two-fold smoothing: We first estimate a local tangent plane at each 3D point and then reconstruct each 3D point by weighted averaging of its projections on multiple tangent planes. We also provide the theoretical analysis for the surface normal estimation and achieve a tighter bound than in a previous work. We validate the empirical performance on the dataset of ShapeNetCore and show that weighted multi-projection outperforms its competitors in all nine classes.
\end{abstract}


\section{Introduction}
With the rapid development of 3D sensing techniques and image-based 3D reconstruction techniques, 3D points are increasingly used in robotics, autonomous driving, augmented reality, computer-aided shape design and many other practical tasks involving objects or environment reconstruction \cite{pcl, pc_tutorial,sample_surface}. 

3D point clouds are usually obtained from two approaches: 3D scanning devices and image-based 3D reconstruction~\cite{sensor,image_pc}. Both approaches introduce noise. For example, 3D scanners suffer from measurement noise, especially around edges or corners of objects; image-based 3D reconstruction often fails to manage matching ambiguities or image imperfection \cite{machine_noise}. Noise not only makes the reconstructed surface lose fine details and deforms the underlying manifold structure but it also impairs subsequent geometry processing and computation, such as compression, segmentation, registration and recognition~\cite{siheng}. However, 3D point cloud denoising is challenging; 1D time series and 2D images support on grids, a 3D point cloud is usually a set of unordered points scattered in the 3D space without any connectivity or topology information. 

\begin{figure}
        \centering
        \begin{subfigure}[b]{0.225\textwidth}
            \centering
            \includegraphics[width=1.3in]{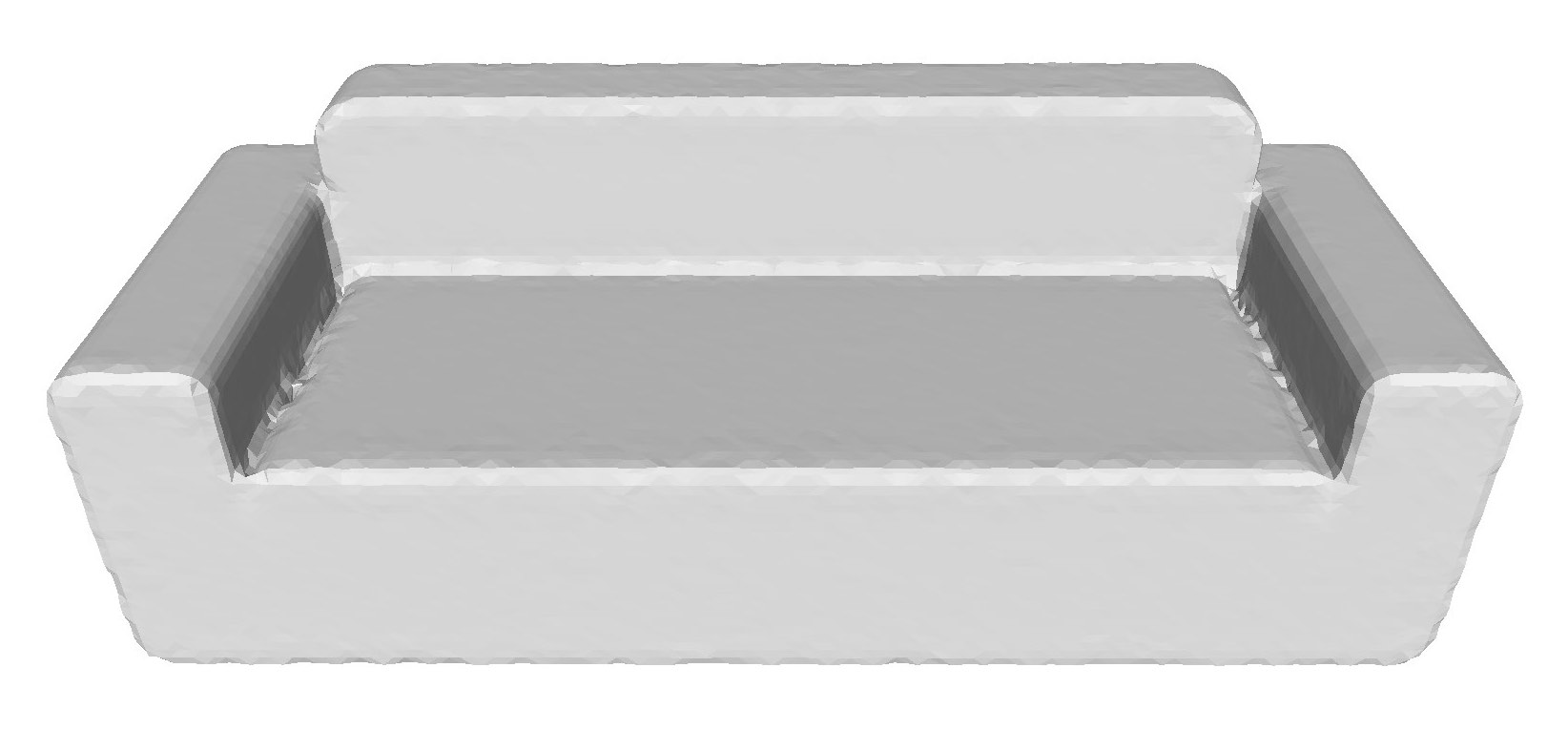}
            \caption[]%
            {{\small Noise-free point cloud.}}    
        \end{subfigure}
        \hfill
        \begin{subfigure}[b]{0.225\textwidth}  
            \centering 
            \includegraphics[width=1.3in]{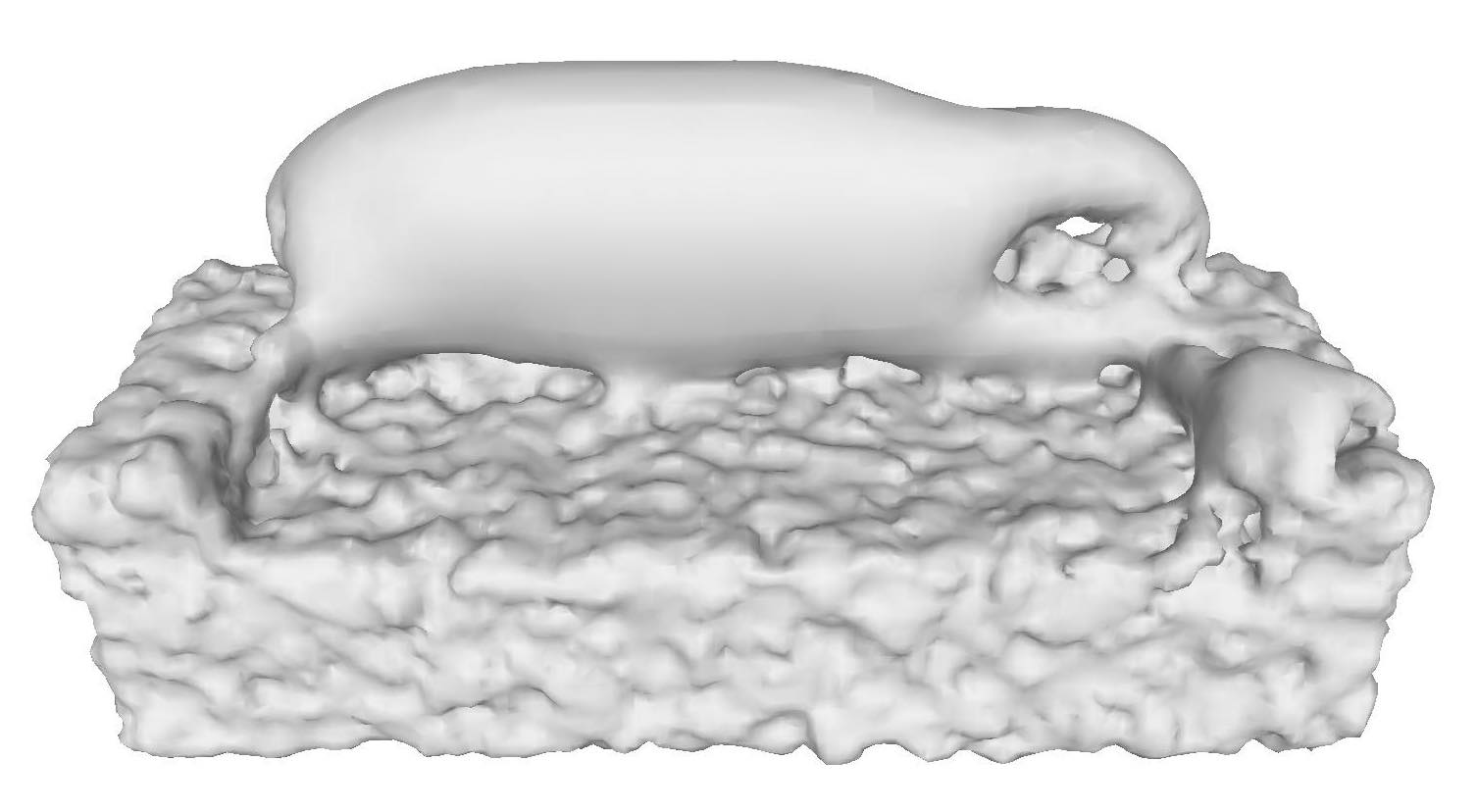}
            \caption[]%
            {{\small Noisy point cloud.}}    
        \end{subfigure}
        \vskip 0.01cm
        \begin{subfigure}[b]{0.225\textwidth}   
            \centering 
            \includegraphics[width=1.3in]{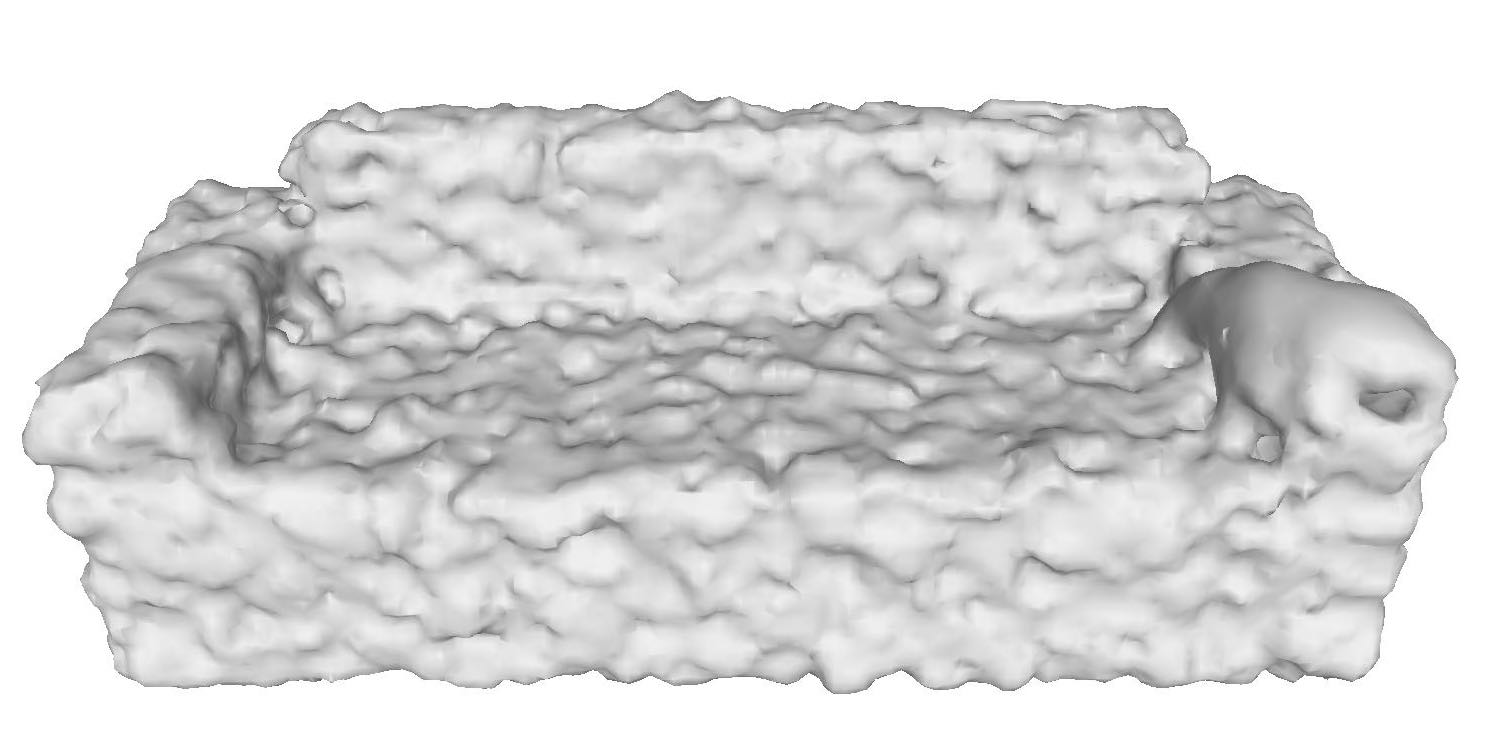}
            \caption[]%
            {{\small Bilateral filter.}}    
        \end{subfigure}
        \quad
        \begin{subfigure}[b]{0.225\textwidth}   
            \centering 
            \includegraphics[width=1.3in]{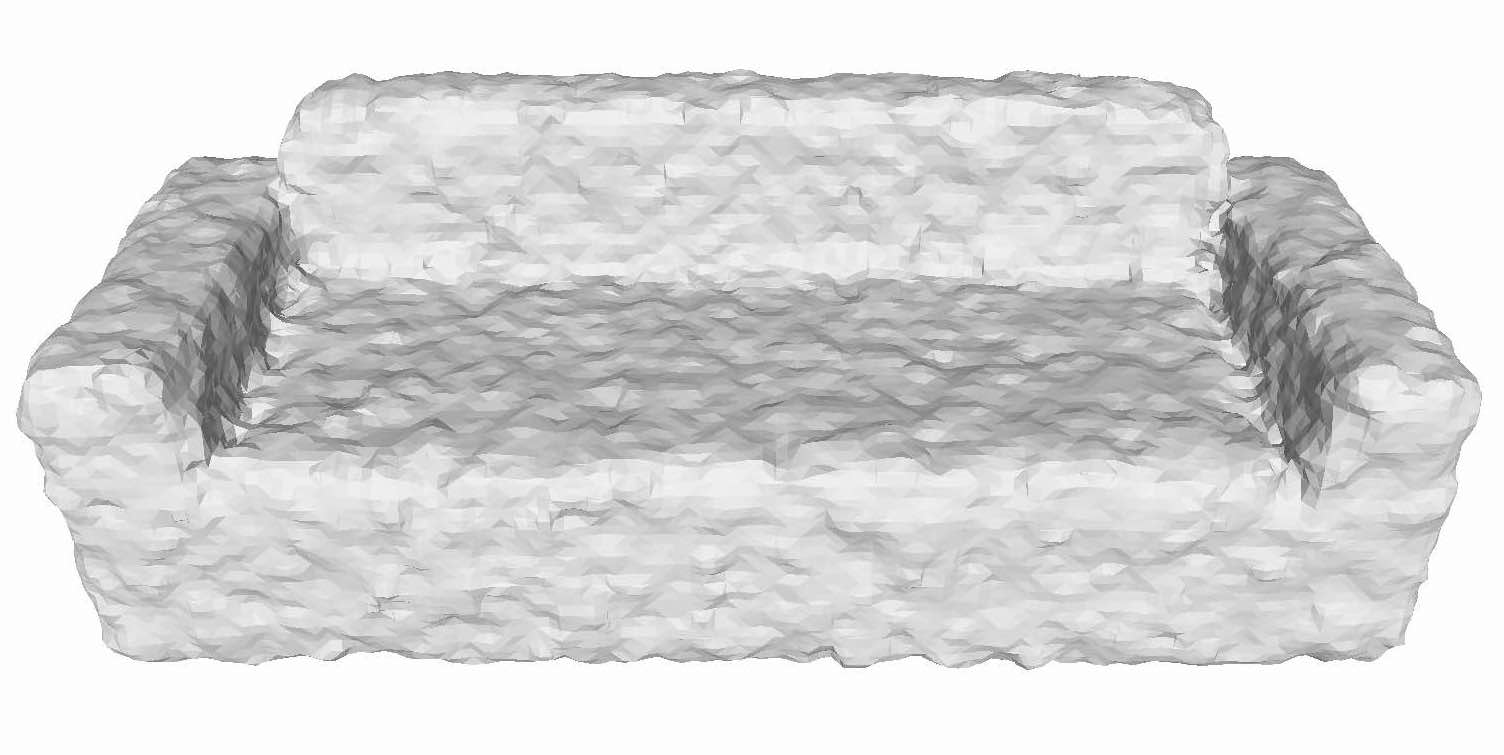}
            \caption[]%
            {{\small Graph-based filter.}}    
        \end{subfigure}
        %\vskip\baselineskip
        \vskip 0.01cm
        \begin{subfigure}[b]{0.225\textwidth}   
            \centering 
            \includegraphics[width=1.3in]{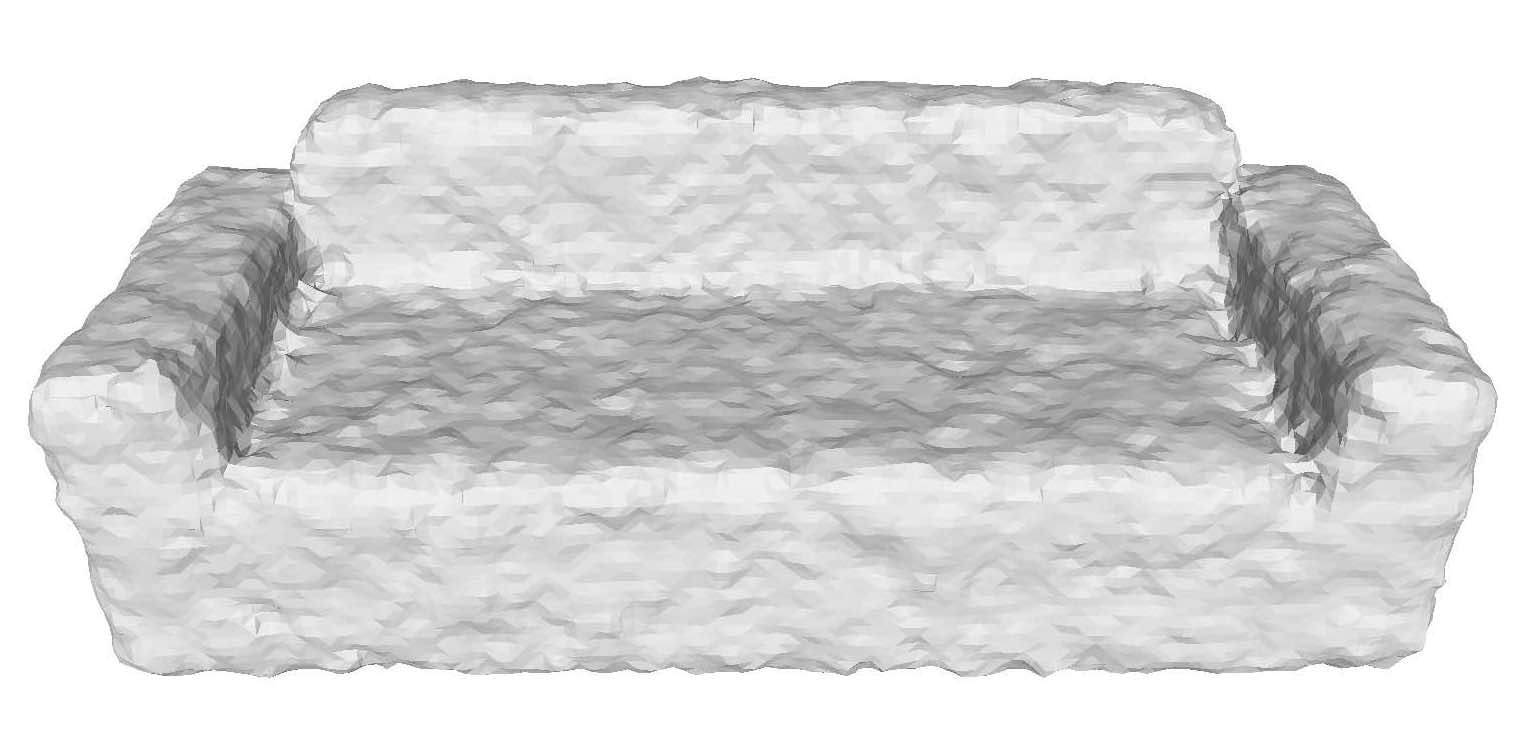}
            \caption[]%
            {{\small PDE-based filter.}}    
        \end{subfigure}
        \quad
        \begin{subfigure}[b]{0.225\textwidth}   
            \centering 
            \includegraphics[width=1.3in]{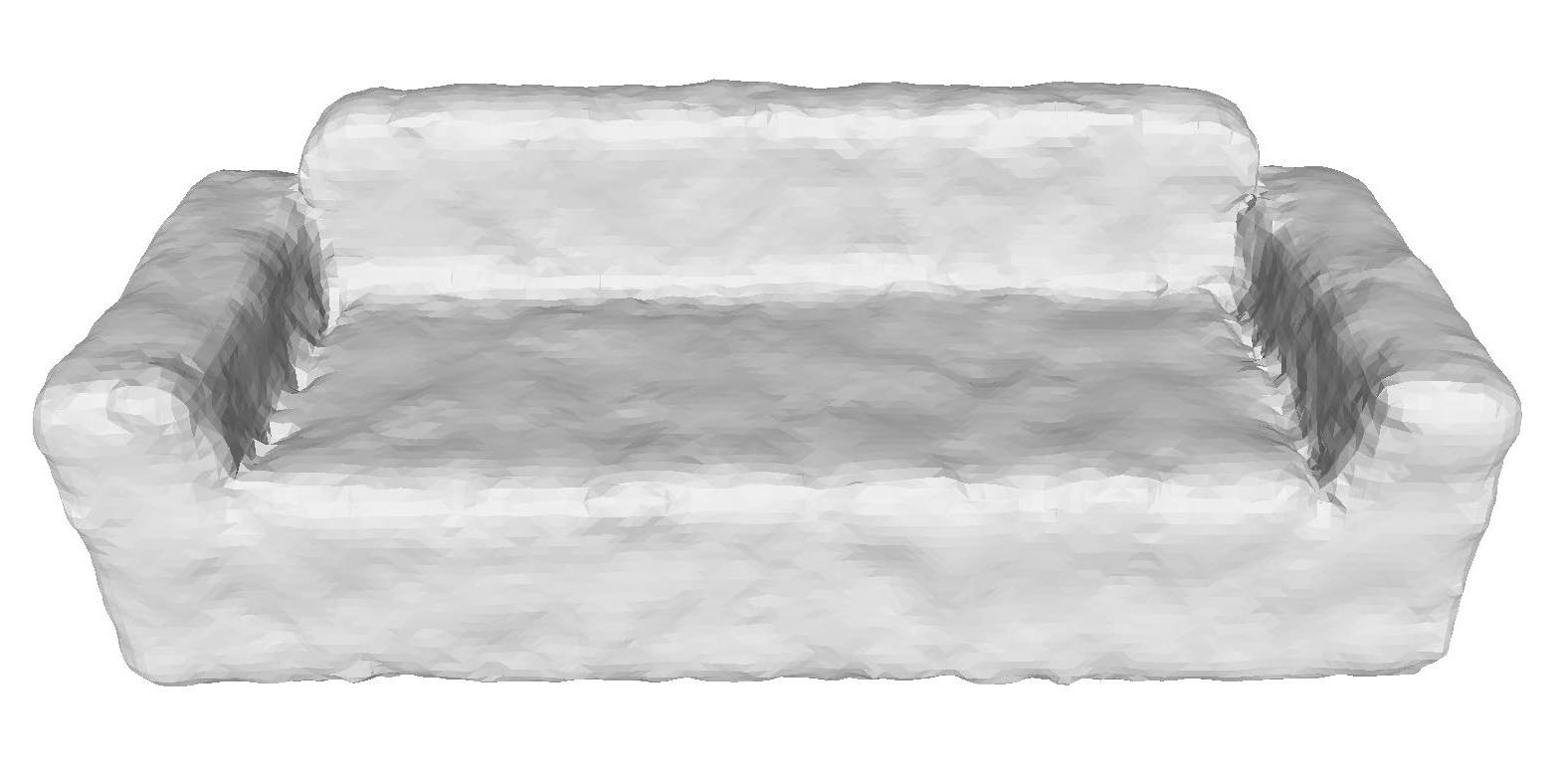}
            \caption[]%
            {{\small Proposed WMP.}}    
        \end{subfigure}
        \caption[]
        {\small The proposed WMP produces a cleaner and smoother mesh reconstruction compared to its competitors. Plots (a) and (b) show the mesh reconstructed from noiseless point cloud and noisy point cloud, respectively. Plots (c)-(f) show the meshes reconstructed from the denoised point cloud produced by different algorithms.}
        \vspace{-5mm}
        \label{mesh_sofa}
    \end{figure}

To tackle the problem of 3D point cloud denoising, we propose a novel algorithm, called weighted multi-projection (WMP), which includes three modules, graph construction, tangent plane estimation, and reconstruction from multiple projections. We first construct a graph based on the 3D coordinates of points; this graph captures the local and global geometric structures of the underlying manifold. Based on the graph, we then estimate the local tangent planes for all the points, which approximates the surface from which the points are sampled. After that, a multi-projection procedure is followed by weighted averaging. The intuition behind this algorithm is to use the tangent planes to remove the orthogonal component of noise. We further show a theoretical bound for the estimation error of surface normals in the 2D case, which is proved to be a tighter bound comparing with~\cite{normal_bound} and show that the multiple-projection strategy is better than the one-time projection strategy. To validate the empirical performance, we also test the proposed algorithm on the real dataset of ShapeNetCore. The experimental results show that our algorithm outperforms other algorithms in all nine categories. 

{\bf Related Work.}
3D point cloud denoising has been tackled by various approaches: mesh-based denoising, bilateral-filter-based denoising, partial-differential-equation-based denoising and graph-based denoising. Most mesh-based denoising algorithms use meshes constructed from 3D points as the input~\cite{mesh1}, \cite{mesh2}. Since mesh construction algorithms are time-consuming and are sensitive to noise~\cite{denoise_bf}, researchers have explored algorithms to directly deal with 3D point clouds. The authors in \cite{denoise_bf} adapt the bilateral filter (BF) from mesh denoising~\cite{bf_2d}, \cite{bf_mesh} and implement it to denoise point clouds. However, mesh-based algorithms are not fully transferable and causes unavoidable shrinkage and deformation~\cite{denoise_nld}. Partial differential equations (PDE) are widely used in image and mesh denoising~\cite{classify2}; they have been extended to 3D point cloud denoising~\cite{denoise_pde1}, \cite{denoise_pde2}; however, PDE-based denoising algorithm cause local over-smoothing \cite{classify2}.  Graph-based denoising (GBD) algorithms have received increasing attention because the Laplace-Beltrami operator of manifolds can be approximated by the graph Laplacian~\cite{graph_manifold}; however, a constructed graph is also noisy and cannot reflect the true manifold. In practice, graph signal denoising algorithms cause clustering and deformation issues~\cite{graph_analysis}. 

Note that we also use the graph structure to capture the local geometric structure. Instead of using the graph to denoise directly, like those classical graph-based algorithms, we use the graph structure to estimate the tangent planes and use the estimated tangent planes to denoise. This two-fold smoothing process ease the clustering and deformation issues.

\section{3D point cloud denoising Algorithm}\label{method} 
The proposed point cloud denoising algorithm consists of three modules: graph construction, tangent plane estimation and weighted projection. (1) We first construct a graph based on 3D coordinates of a point cloud. The graph is used to capture the geometric structure of the point cloud. (2) Based on this graph structure, we estimate a tangent plane at each point to locally approximate the underlying manifold. (3) We then project each point onto its own tangent plane and its neighboring points' tangent planes, called the multi-projection. We finally obtain the denoised points by weighted averaging of the coordinates of all the projections. The key component of the proposed algorithm is the tangent plane.

{\bf Problem Setting.} Let $\mathcal{S} = \{\widetilde{\mathbf{p}}_i \in \mathbb{R}^3 \;|\; i = 1, ..., N\}$ be a noisy 3D point cloud, where $N$ is the total number of points in $\mathcal{S}$, and each element $\widetilde{\mathbf{p}}_i = [x_i, \, y_i, \, z_i]^T$ is the noisy coordinate vector of the $i$th point. Note that $\widetilde{\mathbf{p}}_i = \mathbf{p}_i + \mathbf{n}_i$, where $\mathbf{p}_i$ is the noiseless coordinate vector of the $i$th point and $\mathbf{n}_i$ is the noise vector attached to the $i$th point. Since 3D point clouds are sampled from smooth surfaces, they are essentially 2D manifolds embedded in the 3D space. These smooth surfaces can be locally approximated by tangent planes. We thus use the tangent planes to remove noise.
\begin{theorem}
Let $T$ be the tangent plane associated with the noiseless point $\mathbf{p}$; $\mathbf{t}$ be the projection of a noisy point $\widetilde{\mathbf{p}}$ on $T$. Then,
$
\left\| \mathbf{t} - \mathbf{p} \right\|_2^2 \leq \left\| \widetilde{\mathbf{p}} - \mathbf{p} \right\|_2^2.
$
\end{theorem}
The proof is obtained from the projection theorem~\cite{jelena}. The projection on the tangent plane  removes the orthogonal component of noise. Inspired by this, we aim to estimate tangent planes from noisy points and project noisy points on the estimated tangent planes to obtain denoised points.

{\bf Graph Construction.} To capture the local and global geometric structure of a point cloud, we construct a graph $\mathcal{G}$ whose nodes are 3D points and edges are proximities of points. Here we build an $\epsilon$-nearest neighbor ($\epsilon$-NN) graph. The $i$th and $j$th points are connected by an edge when the Euclidean distance between $\widetilde{\mathbf{p}}_i$ and $\widetilde{\mathbf{p}}_j$ is smaller than the pre-defined $\epsilon$. We then construct a weighted adjacent matrix $W  \in \mathbb{R}^{N \times N}$, whose elements are defined as
\begin{equation*}
W_{ij}=
\begin{cases}
\frac{1}{Z_i}\exp(-\frac{\lVert \widetilde{\mathbf{p}}_i - \widetilde{\mathbf{p}}_j \rVert_2^2}{2\sigma^2}) & \text{if } \lVert \widetilde{\mathbf{p}}_i - \widetilde{\mathbf{p}}_j \rVert_2^2 \leq \epsilon, \\
0& \text{otherwise},
\end{cases}
\end{equation*}
where the normalization term is $Z_i = \sum_{j=1}^N \exp(-{\lVert \widetilde{\mathbf{p}}_i - \widetilde{\mathbf{p}}_j \rVert_2^2}/{\left(2\sigma^2\right)}),$
with hyperparameters $\sigma$ and $\epsilon$ controlling the decay rate. Let $\mathcal{N}_i$ be the neighboring set for the $i$th point, containing all the neighboring points that connect to the $i$th point, $\mathcal{N}_i = \{j\;|\;W_{ij} \neq 0 \}$.

{\bf Tangent Plane Estimation.} Based on the constructed graph, we estimate the local tangent plane at each 3D point. Let $T_i = \{\mathbf{t} \in \mathbb{R}^3 | \mathbf{a}_i^T \mathbf{t} = c_i\}\,$ be the tangent plane at the $i$th point, where $\mathbf{a}_i \in \mathbb{R}^3$ is the normal vector, $c_i \in \mathbb{R}$ is the intercept of plane $T_i$ and $\mathbf{t}$ is the point on the plane $T_i$. Following the work in \cite{surface_normal}, we can obtain $\mathbf{a}_i$ and $c_i$ by minimizing the sum of weighted square distances from all the points in $\mathcal{N}_i$ to $T_i$,

\begin{equation*}
\begin{aligned}
	\mathbf{a}_i\;,\;c_i &= \mathop{\arg\min}_{\mathbf{a} , c}\sum_{j \in \mathcal{N}_i} W_{ij}\big(\mathbf{a}^T\widetilde{\mathbf{p}}_j - c\big)^2,\\
    \text{s.t.} & \quad  \lVert\mathbf{a}_i\rVert_2^2 = 1.
\end{aligned}
\end{equation*}
We can show that the normal vector $\mathbf{a}_i$ is simply the normalized eigenvector corresponding to the smallest eigenvalue of the weighted covariance matrix $\widetilde{M}_i \in \mathbb{R}^{3 \times 3}$ defined as follows,
\begin{equation}
\label{eq:covariance_matrx}
	\widetilde{M}_i = \sum_{j \in \mathcal{N}_i} W_{ij}\widetilde{{\mathbf{p}}}_j\widetilde{{\mathbf{p}}}_j^T - \overline{\mathbf{p}}_i\overline{\mathbf{p}}_i^T,
\end{equation}
with
$\overline{\mathbf{p}}_i = \sum_{j \in \mathcal{N}_i} W_{ij}\widetilde{{\mathbf{p}}}_j.$
Once we obtain the normal vector $\mathbf{a}_i$, the intercept is calculated as $c_i = \mathbf{a}_i^T\overline{\mathbf{p}}_i$.

{\bf Weighted Multi-Projection.} As shown earlier, we can use the projection to remove noise. Let $\widetilde{T}_i$ be the estimated tangent plane associated with the $i$th point $\widetilde{\mathbf{p}}_i$ and $\mathbf{t}$ be the point on the plane. We define the denoised point $\widehat{\mathbf{p}}_i$ as follows:
\begin{equation}
\label{eq:one_time_projection}
\widehat{\mathbf{p}}_i = \mathop{\arg\min}_{\mathbf{t} \in \widetilde{T}_i} \lVert \widetilde{\mathbf{p}}_i - \mathbf{t} \rVert_2^2,
\end{equation}
which is called the one-time projection.
Based on the geometry, we know that $\widehat{\mathbf{p}}_i$ is the projection of point $\widetilde{\mathbf{p}}_i$ onto the estimated tangent plane $\widetilde{T}_i$ and we have
$$\widehat{\mathbf{p}}_i = \widetilde{\mathbf{p}}_i - \mathbf{a}_i^T\widetilde{\mathbf{p}}_i \mathbf{a}_i + c_i \mathbf{a}_i.$$

Note that the tangent plane $\widetilde{T}_i$ used in~\eqref{eq:one_time_projection} is not the ground-truth tangent plane $T_i$ and the neighboring points are not guaranteed to be on a same plane due to the sampling density and local curvature, which could influence the denoising performance. Since the neighboring points share similar geometry information \cite{denoise_nld}, the tangent planes of neighboring points are inherently similar; thus, we can project each point to its neighbors' tangent planes and average the multiple projections to obtain the denoised point. For the $i$th point $\widetilde{\mathbf{p}}_i$, we project it onto all of its neighbors' tangent planes, 
\begin{equation*}\label{multi_project1}
\mathbf{t}^{(i)}_j = \mathop{\arg\min}_{\mathbf{t} \in \widetilde{T}_j} \lVert \widetilde{\mathbf{p}}_i - \mathbf{t} \rVert_2^2, \quad j \in \mathcal{N}_i.
\end{equation*}
We then obtain the denoised point $\widehat{\mathbf{p}}_i$ by minimizing the weighted sum of distance from $\widehat{\mathbf{p}}_i$ to $\mathbf{t}_j$ as,
\begin{equation}
\label{eq:multi_projection}
\widehat{\mathbf{p}}_i = \mathop{\arg\min}_{\mathbf{p}}\sum_{j \in \mathcal{N}_i}W_{ij} \lVert \mathbf{p} - \mathbf{t}^{(i)}_j \rVert_2^2.
\end{equation}
The solutions is simply $\widehat{\mathbf{p}}_i = \sum_{j \in \mathcal{N}_i}W_{ij}\mathbf{t}_j^{(i)}.$ The proposed WMP algorithm for a given noisy point $\widetilde{\mathbf{p}}_i \in \mathcal{S}$ is summarized up in Algorithm \ref{wmp}.
\begin{algorithm}[htbp]
\caption{Weighed multi-projection (WMP)}
\begin{algorithmic} \label{wmp}
\REQUIRE A noisy point cloud $\mathcal{S} = \{ \widetilde{\mathbf{p}}_i \}$,
\\
~~~~~~~~A hyperparameter $\epsilon$-NN graph construction in $\epsilon$ 
\ENSURE A denoised point cloud $\widehat{\mathcal{S}}$
\STATE Construct $\epsilon$-NN graph and calculate weight matrix $W$
\FOR{$\widetilde{\mathbf{p}}_i \in \mathcal{S}$}
\STATE $\mathcal{N}_i \gets$ neighbors of $\widetilde{\mathbf{p}}_i$ defined by $\epsilon$ 
\STATE $\overline{\mathbf{p}}_i \gets \sum_{j \in \mathcal{N}_i} W_{ij} \widetilde{\mathbf{p}}_j$
\STATE $\widetilde{M}_i \gets \sum_{j \in \mathcal{N}_i} W_{ij}\widetilde{{\mathbf{p}}}_i\widetilde{{\mathbf{p}}}_j^T - \overline{\mathbf{p}}_i\overline{\mathbf{p}}_i^T$
\STATE $\mathbf{a}_i \gets \arg\min \mathbf{a}^T \widetilde{M}_j \mathbf{a}, \quad \lVert \,\mathbf{a}\,\rVert = 1$
\STATE $c_i \gets \mathbf{a}_j^T \overline{\mathbf{p}}_j$, 
\ENDFOR
\FOR{$\widetilde{\mathbf{p}}_i \in \mathcal{S}$}
\STATE $\mathbf{t}_j^{(i)} \gets \widetilde{\mathbf{p}}_i - \mathbf{a}_j^T\widetilde{\mathbf{p}}_i \mathbf{a}_j + c_j \mathbf{a}_j$
\STATE $\widehat{\mathbf{p}}_i \gets \sum_{j \in \mathcal{N}_i}W_{ij}\mathbf{t}_j^{(i)}$
\ENDFOR
\end{algorithmic}
\end{algorithm}

{\bf Computational Cost.}
For a point cloud with $N$ points, the computational complexities for graph construction and tangent plane estimation are $O(N\log N)$. The projection and weighted summation take $O(kN)$ time, where $k$ is the average neighbors of each point decided by $\epsilon$ in the $\epsilon$-NN graph. In total, our computational complexity is $O(N(\log N + k))$.

% \begin{figure}[htbp]
% \centerline{\includegraphics[width=3.2in]{reference.png}}
% \caption{Projection}\label{reference}
% \end{figure}

\vspace{-3mm}
\section{Theoretical Analysis}\label{math}
In this section, we show the theoretical performance of the proposed WMP. For the simplicity, We just show a 2D case, where 2D points lies along a semicircle. We could generalize the results to 3D point clouds by replacing a semicircle to a sphere, which is left for future works.

\begin{figure}[htbp]
\centerline{\includegraphics[width=2.4in]{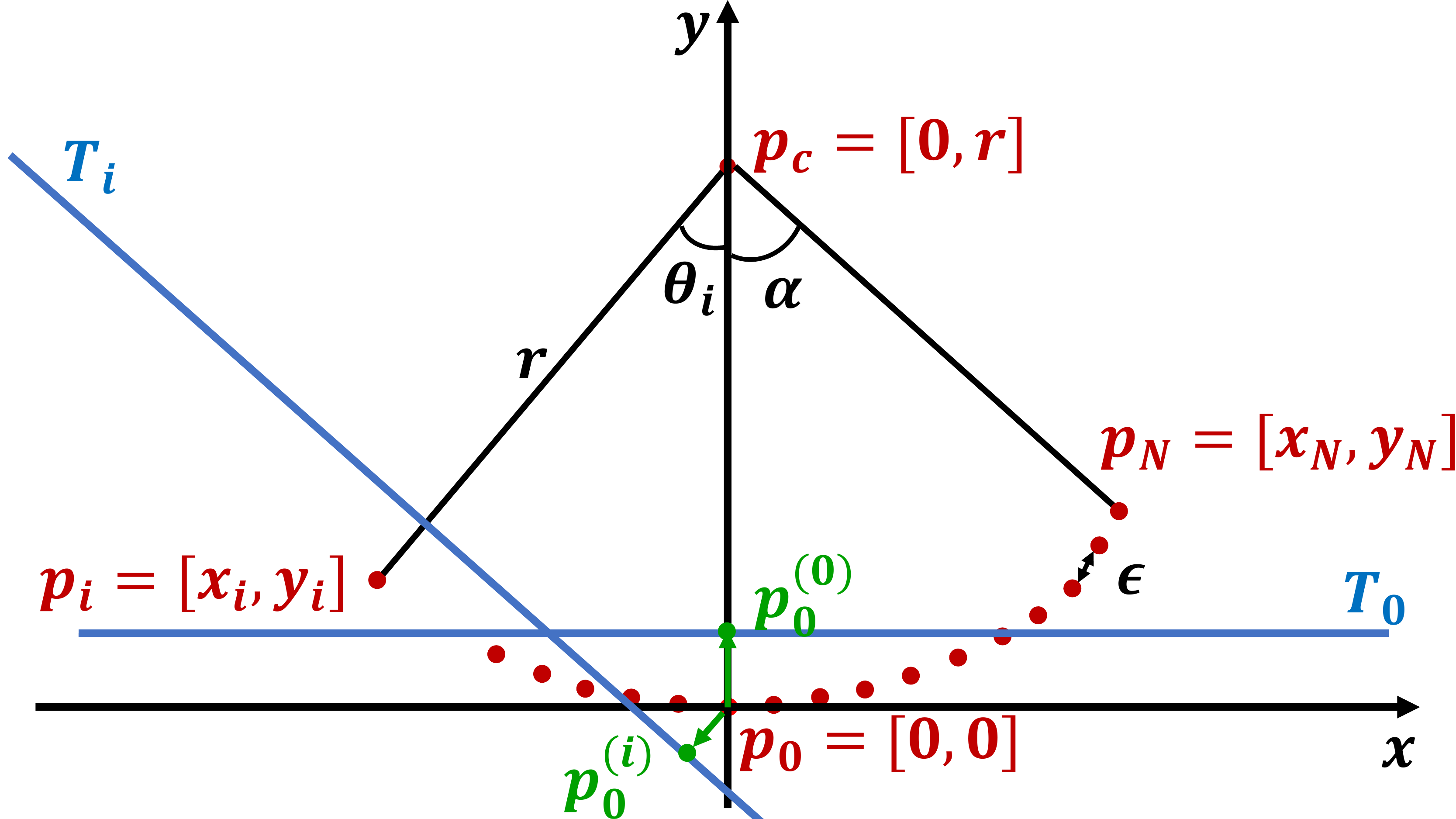}}
\caption{The red dots represent the points sampled from a semicircle centered at $\mathbf{p}_c = [0\, , \,r]$ with radius $r$. The coordinate of point $p_i$ is represented as $[x_i\, , \, y_i],\, -N \leq i \leq N$; $\mathbf{p}_0 = [0\, ,\, 0]$ is the origin point. The blue lines $T_0$ and $T_i$ denote the tangent lines for point $\mathbf{p}_0$ and $\mathbf{p}_i$, respectively. The tangent lines are reconstructed based on their neighboring points. The length of the double arrows line is $\epsilon$, representing the distance between neighboring points. The green point on the tangent lines $\mathbf{p}_0^{(i)}$ represents the projection of point $\mathbf{p}_0$ onto tangent line $T_i$. We use the angel $\theta_i$ to locate the position of points.}\label{circle_curve}
\end{figure}

Let a set of $N$ 2D points be uniformly sampled from a semicircle centered at the center $[0\,,\, r]$ with radius $r$. The $i$th point in a noiseless point cloud is
$$
\mathbf{p}_i \ = \ \begin{bmatrix}
x_i \\ y_i
\end{bmatrix}
= \begin{bmatrix}
r\sin \theta_i \\ r - r\cos \theta_i
\end{bmatrix},
$$
and the corresponding noisy point is
$$\widetilde{\mathbf{p}}_i \ = \ \begin{bmatrix}
\widetilde{x}_i \\ \widetilde{y}_i
\end{bmatrix}
= \begin{bmatrix}
r\sin \theta_i + n_{x,i}
\\ 
r - r\cos \theta_i + n_{y,i}
\end{bmatrix},
$$
where $\theta_i = \epsilon i / r$ is the interval angle with $\epsilon$ a constant representing the interval distance between two points, $n_{x , i}$ and $n_{y,i}$ are the $x$ and $y$ components of noise and $i \in [-N\, , \, N]$. Following this, we have two properties:
\begin{enumerate}
\item The derivative at origin point $\mathbf{p}_0 = [0\, , \, 0]$ is 0 and the normal vector at point $P$ is $[0\, ,\, 1]^T$.
\item The curvature of all the sampled points is $\kappa = 1/r$.
\end{enumerate}

% \subsection{Modeling}\label{model}
% We assume all the points are sampled from a curve in $\mathbb{R}^2$, which can be represented as,
% \begin{equation}\label{2dcurve} 
% y = \frac{\kappa}{2}x^2, \quad \kappa \geq 0,
% \end{equation}
% and let $O$ be the origin point $[0~,~0]^T$.
% We choose function (\ref{2dcurve}) for the following properties,
% \begin{enumerate}
% 	\item The derivative at point $O$ is $0$ and normal vector at this point is $[0~,~1]^T$. 
% 	\item The curvature at point $O$ is $\kappa$. 
% \end{enumerate}

% Let $\mathbf{p}_i = [x_i~,~y_i]^T$ for $0 \leq i \leq 2k $ be the points uniformly sampled from the curve, and we have $y_i = \frac{\kappa}{2}x_i^2$ when the sampled 2D point cloud is noise free. Assume $x_i$ follows a uniform distribution in the interval of $[-r , r]$ and $x_{i + 1} - x_i = \epsilon, ~0 \leq i \leq (2k - 1)$. For the noisy 2D point cloud, we have $y_{n,i} = \frac{\kappa}{2}x_i^2 + n_i$, where $n_i$ is the noise term taking the value from $[-n , n]$. We assume that $x_i$ and $n_i$ are independent and the mean of noise $n_i$ is zero.

Let $M_0, \widetilde{M}_0 \in \mathbb{R}^{2 \times 2}$ be the covariance matrices of the noiseless point $\mathbf{p}_0$, and the noisy point $\widetilde{\mathbf{p}}_0$ respectively, Following the definition in~\eqref{eq:covariance_matrx}, we have
\begin{equation*}
	\begin{aligned}
	M_0 & = \frac{1}{(2N + 1)}\left[\begin{matrix}
	\sum_{j = -N}^{N}x_j^2 & \sum_{j = -N}^{N}x_j y_j \\
	\sum_{j = -N}^{N}x_j y_j & \sum_{j = -N}^{N}y_j^2
	\end{matrix}\right] \\
	& = \left[\begin{matrix}
	m_{11} & m_{12} \\
	m_{12} & m_{22}
	\end{matrix}\right],
	\end{aligned}
\end{equation*}

\begin{equation*}
	\begin{aligned}
	\widetilde{M}_0 & = \frac{1}{(2N + 1)}\left[\begin{matrix}
	\sum_{j = -N}^{N}\widetilde{x}_j^2 & \sum_{j = -N}^{N}\widetilde{x}_j \widetilde{y}_j,\\
	\sum_{j = -N}^{N}\widetilde{x}_j \widetilde{y}_j & \sum_{j = -N}^{N}\widetilde{y}_j^2
	\end{matrix}\right] \\
	& = \left[\begin{matrix}
	\widetilde{m}_{11} & \widetilde{m}_{12} \\
	\widetilde{m}_{12} & \widetilde{m}_{22}
	\end{matrix}\right].
	\end{aligned}
\end{equation*}
Let $\mathbf{a}, \widehat{\mathbf{a}}$ be the ground-truth normal vector at point $\mathbf{p}_0$, and the estimated normal vector at point $\widetilde{\mathbf{p}}_0$, respectively. We can bound the error of the surface normal estimation based on Davis-Kahan Theorem~\cite{davis_theorem}.

\begin{theorem}
Let $\theta$ be the angle difference between the ground-truth normal vector $\widetilde{\mathbf{a}}$ and the estimated normal vector $\mathbf{a}$. Then,
\begin{equation*}
|\sin \theta| \leq \frac{|\widetilde{m}_{12}| + |\widetilde{m}_{22} - m_{22}| }{m_{11}}.
\end{equation*}
\end{theorem}
We omit the proof here due to space constraints. The previous work on surface normal estimation~\cite{surface_normal} considers noise is added to $y$ axis and the error bound $\theta$ is
\begin{equation*}
|\theta| \leq \frac{|\widetilde{m}_{12}| + \widetilde{m}_{22}}{m_{11}}.
\end{equation*}
Compared to this bound, we assume that noise is added to both $x$ and $y$ axes and our bound is tighter when the SNR of a noisy signal is higher than $0$.

We now show that the multi-projection~\eqref{eq:multi_projection} is theoretically better than the one-time projection~\eqref{eq:one_time_projection}. 
\begin{theorem}
Let $\mathbf{p}_0 = [0, 0]$ be the origin point.
Let $\widehat{\mathbf{p}}_{\rm o}$ and $\widehat{\mathbf{p}}_{\rm m}$ be the reconstructions of $\mathbf{p}_0$ based on the one-time projection~\eqref{eq:one_time_projection}
and the multi-projection~\eqref{eq:multi_projection}, respectively. Then, we have
\begin{equation*}
\begin{aligned}
e_{\rm o}(L, \kappa) = \left\| \widehat{\mathbf{p}}_{\rm o} - \mathbf{p}_0 \right\|_2 & =  \frac{1}{\kappa}(1 - \frac{\sin (L \kappa)}{L \kappa}), \\
e_{\rm m}(L, \kappa) = \left\| \widehat{\mathbf{p}}_{\rm m} -\mathbf{p}_0 \right\|_2 & =  \frac{1}{\kappa} (\frac{1}{2}\frac{\sin 2(L \kappa)}{2(L \kappa)} + \frac{1}{2} - \frac{\sin^2 (L \kappa)}{(L \kappa)^2}),
\end{aligned}
\end{equation*}
where $L = \epsilon N > 0$ and $\kappa = 1/r >0$, representing the curvature at point $\mathbf{p}_0$.
\end{theorem}

\begin{figure}[htbp]
\centerline{\includegraphics[width=3.2in]{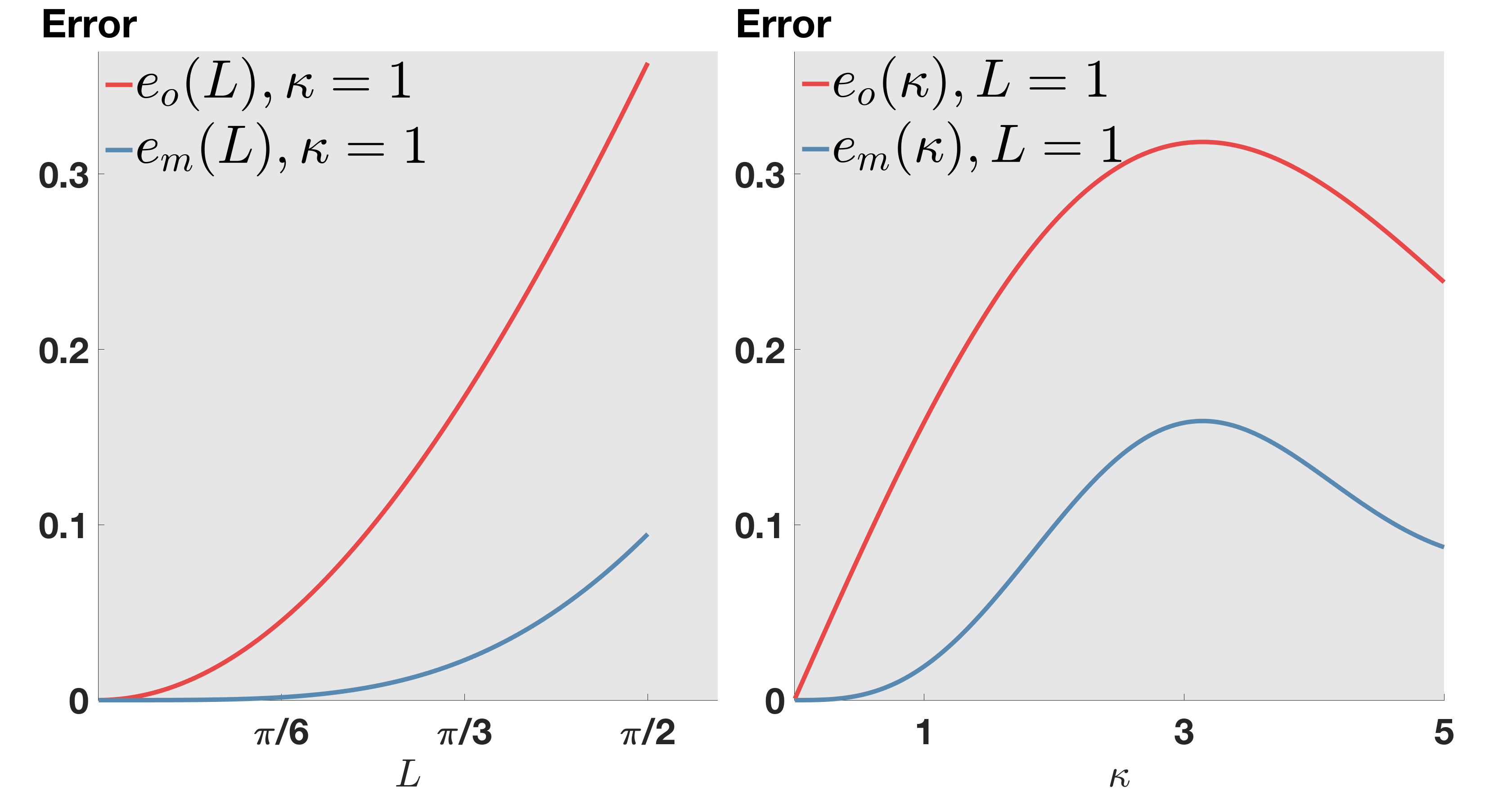}}
\caption{The multi-projection is better than the one-time projection. Left figure shows the errors of two strategies with a fixed curvature $\kappa = 1$. Right figure shows the errors of two strategies with a fixed $L = 1$. In both, multi-projection produces smaller error and changes more slowly near to zero.}
\vspace{-5mm}
\label{one_multi}
\end{figure}

Fig.~\ref{one_multi} compares errors of one-time-projection $e_{\rm o}(L, \kappa)$ and multi-projection $e_{\rm m}(L, \kappa)$ for a fixed $\kappa = 1$ and a fixed $L = 1$, respectively. Both projection strategies produce zero error when $L \rightarrow 0$ or $\kappa \rightarrow 0$, but the proposed multi-projection achieves lower bound for both cases; the multi-projection is more robust than the one-time-projection because the value of $e_{\rm m}(L , \kappa)$ varies much smoother than $e_{\rm o}(L , \kappa)$ for $L \rightarrow 0$ or $\kappa \rightarrow 0$. 
% Fig.~\ref{one_multi} shows that the estimated error is smaller when the curvature $\kappa$ is smaller for same sampling strategy with a constant $L = \epsilon N$; the estimated error is smaller when the distance between neighboring points $\epsilon$ is smaller for same curvature $\kappa$. 
% It may be an instructive sampling strategy for us to update the number of neighboring points according to their curvatures to achieve a smaller estimated error and a better version of denoised point clouds. 

\section{Experimental Results}\label{experiment}
{\bf Dataset.} We evaluate our algorithm on the dataset in ShapeNetCore, which contains 55 manually verified common object categories with more than 50,000 clean 3D mesh models \cite{shapenet}. We first choose nine categories and randomly select 50 items in each category. Then we sample points from the surfaces of these 3D models with Poisson-disk sampling algorithm \cite{possion_disk} and rescale the points into a unit cube centered at the origin. We add Gaussian noise to construct noisy point clouds. We compare the proposed WMP with the state-of-the-art denoising algorithms, including BF algorithm~\cite{denoise_bf}, GDB algorithm~\cite{denoise_graph}, PDE algorithm~\cite{denoise_pde1} and Non-local Denoising (NLD) algorithm~\cite{denoise_nld}. For each of these algorithms, we tune the parameters to produce its own best performance. 

{\bf Results and Analysis.} To quantify the performance of different algorithms, we the following metrics:
signal-to-noise ratio (SNR) defined as,
\begin{equation*}
\text{SNR}(\mathcal{S}_1 , \mathcal{S}_2) = 20 \log \big(\frac{\sum_{\mathbf{p}_i \in \mathcal{S}_1}||\mathbf{\mathbf{p}_i}||_2^2}{\sum_{\mathbf{p}_i \in \mathcal{S}_1, \mathbf{q}_i \in \mathcal{S}_2} ||\mathbf{p}_i - \mathbf{q}_i||_2^2}\big);
\end{equation*} 
mean squared error (MSE) defined as,
\begin{equation*}
\text{MSE}(\mathcal{S}_1 , \mathcal{S}_2) = \frac{1}{N}\sum_{\mathbf{p}_i \in \mathcal{S}_1, \mathbf{q}_i \in \mathcal{S}_2} ||\mathbf{p}_i - \mathbf{q}_i||_2^2;
\end{equation*}
Chamfer distance (CD) defined as,
\begin{eqnarray*}
\text{CD}(\mathcal{S}_1 , \mathcal{S}_2) & = & \frac{1}{N} \Big(\sum_{\mathbf{p}_i\in {\mathcal{S}_1}}\min_{\mathbf{q}_j\in {\mathcal{S}_2}}||\mathbf{p}_i - \mathbf{q}_j||_2^2 
\\
&& + \sum_{\mathbf{q}_i\in {\mathcal{S}_2}}\min_{\mathbf{p}_j\in {\mathcal{S}_1}}||\mathbf{q}_j - \mathbf{p}_i||_2^2 \Big),
\end{eqnarray*}
where $\mathcal{S}_1$ and $\mathcal{S}_2$ represent the denoised point cloud and original point cloud, respectively, $\mathbf{p}_i$ and $\mathbf{q}_i$ represent the coordinates of points and $N$ is the total number of points.

\begin{figure}[htbp]
\vspace{-5mm}
\centerline{\includegraphics[width=3.6in]{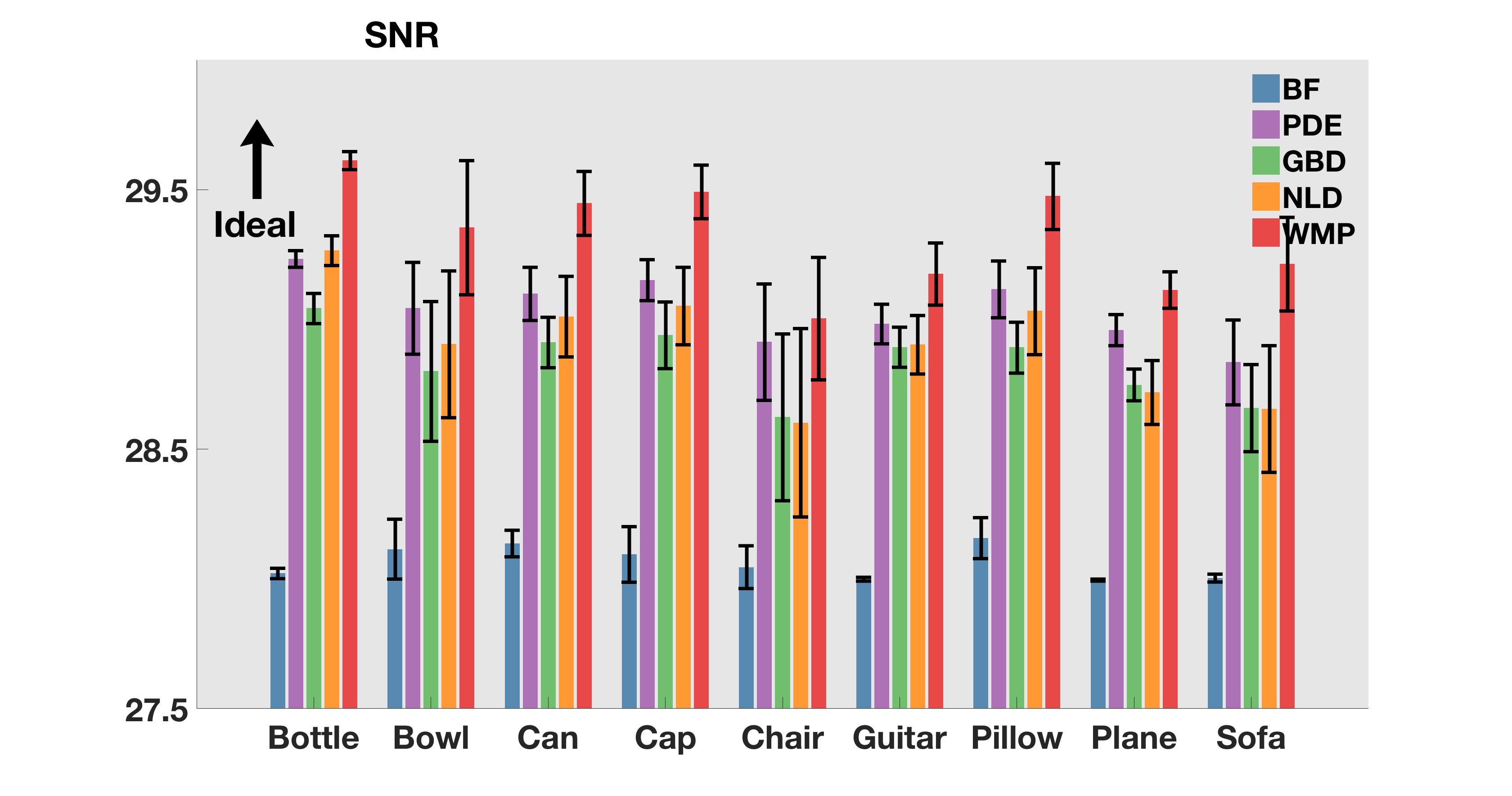}}
\caption{The proposed WMP (in red) outperforms its competitors in terms of SNR. The denoised point clouds produced by WMP contain higher original signal intensity compared to the denoised point clouds produced by other algorithms.}
\vspace{-5mm}
\label{snr_mean}
\end{figure}

\begin{figure}[htbp]
\vspace{-5mm}
\centerline{\includegraphics[width=3.6in]{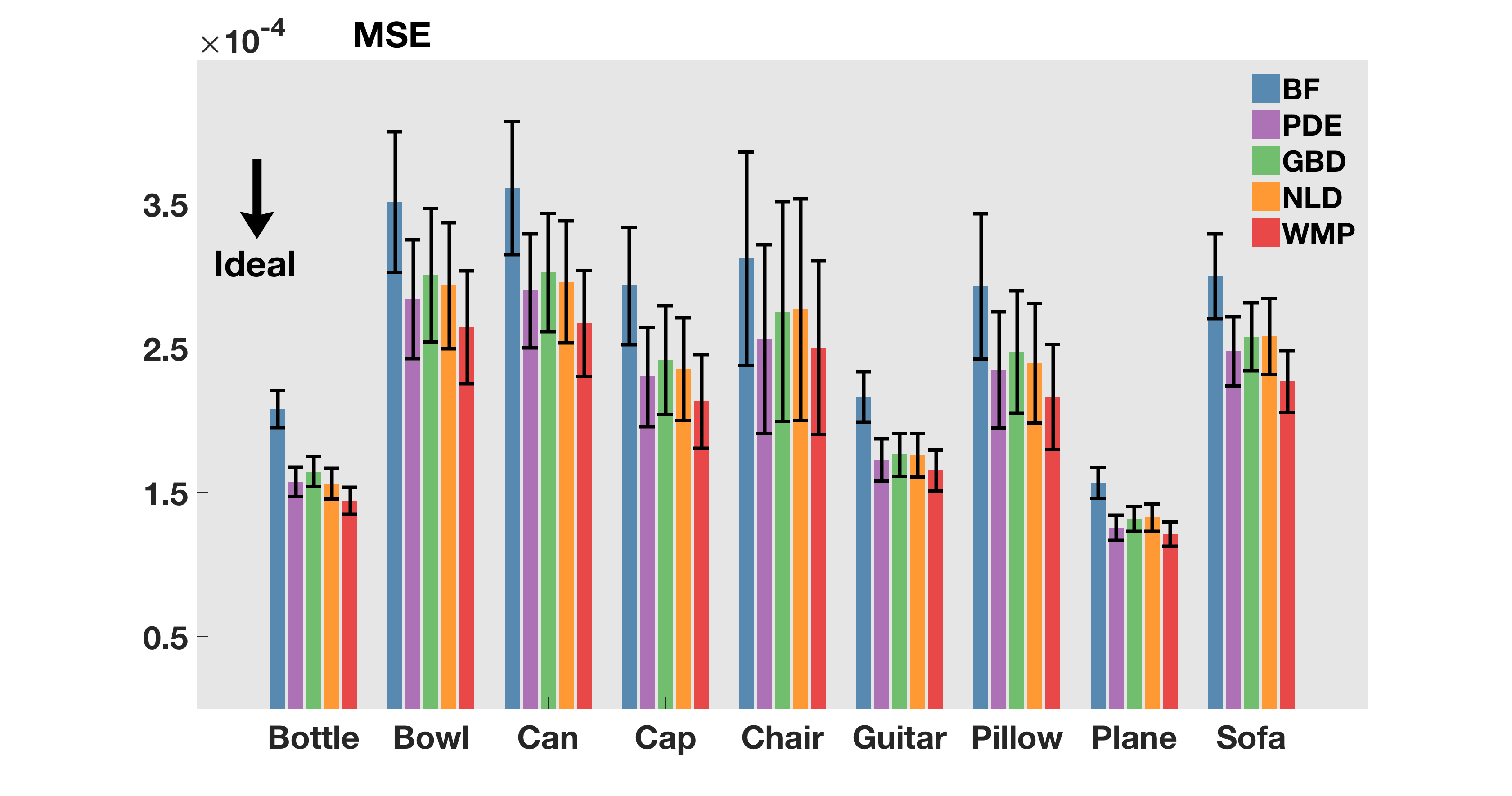}}
\caption{The proposed WMP (in red) outperforms its competitors in terms of MSE. The denoised point clouds produced by WMP stay point-wise closer to the noiseless point clouds compared to the denoised point clouds produced by other algorithms.}
\vspace{-5mm}
\label{mse_mean}
\end{figure}

\begin{figure}[htbp]
\vspace{-5mm}
\centerline{\includegraphics[width=3.6in]{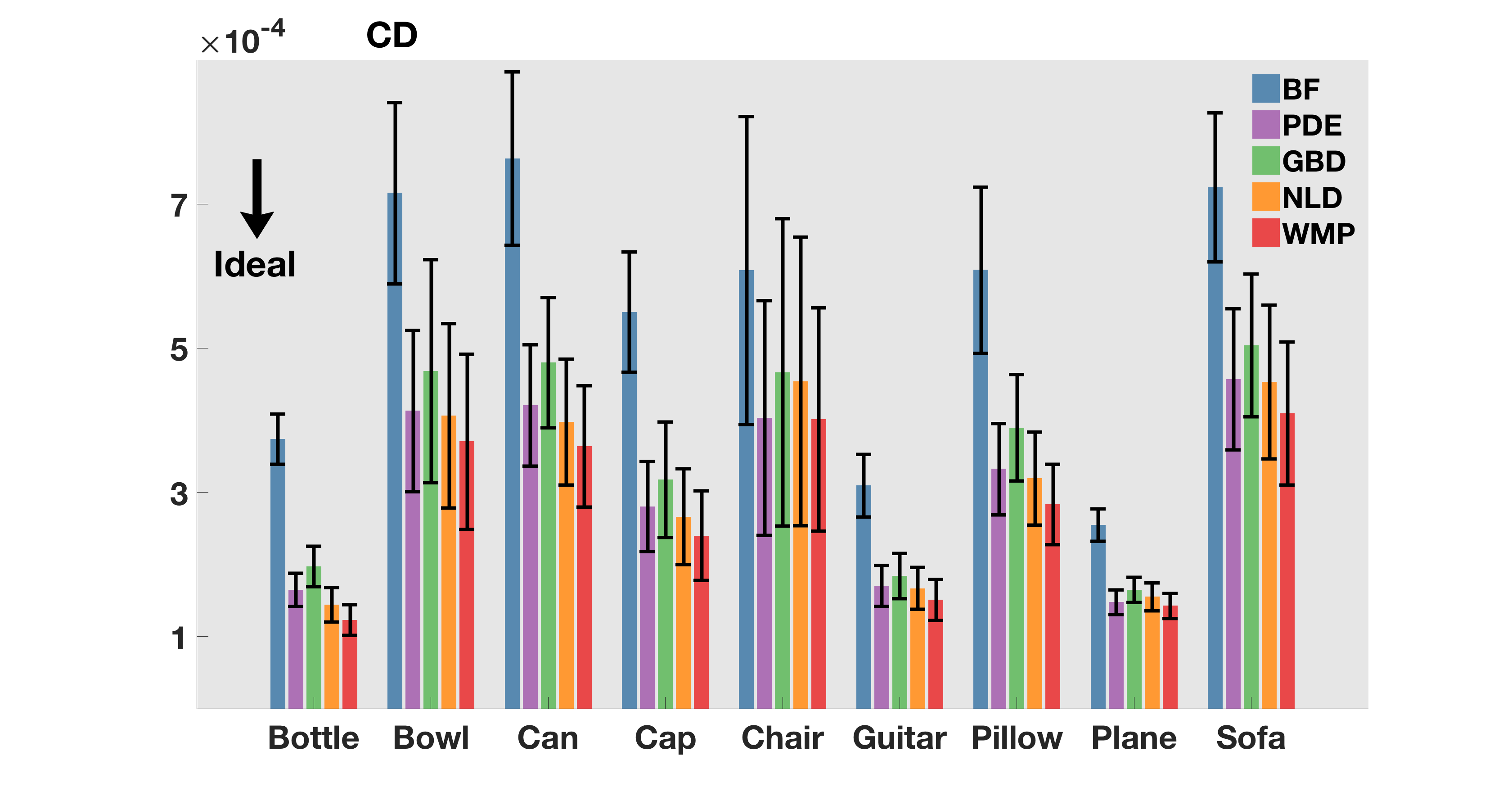}}
\caption{The proposed WMP (in red) outperforms its competitors in terms of CD. The denoised point clouds produced by WMP are closer to the original point clouds or the manifolds point clouds are lying on compared to the denoised point clouds produced by other algorithms.}
\vspace{-5mm}
\label{cd_mean}
\end{figure}

Figs. ~\ref{snr_mean}, \ref{mse_mean} and \ref{cd_mean} show the denoising performances in nine categories evaluated by SNR, MSE and CD (mean and standard deviation for 50 items in each category), respectively. We see that the proposed WMP outperforms its competitors in all nine categories; all algorithms perform slightly worse on chair and sofa datasets, because chair and sofa contain more sharp edges compared with other categories. Fig.~\ref{mesh_sofa} visualizes the denoising performances of a noisy sofa from different algorithms. We construct meshes from a noisy point cloud and its denoised versions by Poisson surface reconstruction algorithm with same parameters in MeshLab \cite{meshlab}. We see that the proposed algorithm outperforms its competitors in all categories of point clouds by all evaluation metrics (SNR, MSE and CD). 

% \begin{table}[ht]
% \caption{Results: SNR}
%   \begin{center}
%   \begin{tabular}{|c|c|c|c|c|c|}
%   \hline
%     Model & BF & PDE & GRAPH & NLD & WMP\\
%     % 28.0444   28.9127   28.6230   28.6008   29.0029
%     \hline
%     Bottle & 28.0206 & 29.2325 & 29.0424 & 29.2652 & \textbf{29.6120
% } \\
%     \hline
%     Bowl & 28.1136 & 29.0432 & 28.7998 & 28.9038 & \textbf{29.3542} \\
%     \hline
%     Can & 28.1367 &  29.0986 &  28.9106  & 29.0106 &  \textbf{29.4475} \\
%     \hline
%     Cap & 28.0942 &  29.1508 &  28.9392  & 29.0514 &  \textbf{29.4913} \\
%     \hline
%     Chair & 28.0444 & 28.9127 & 28.6230 & 28.6008 & \textbf{29.0029} \\
%     \hline
%     Guitar & 27.9993 &  28.9823  & 28.8930 &  28.9031  & \textbf{29.1750} \\
%     \hline
%     Pillow & 28.1571  & 29.1162 &  28.8916  & 29.0322  & \textbf{29.4749}\\
%     \hline
%     Plane & 27.9949  & 28.9589 & 28.7474 & 28.7189 & \textbf{29.1134} \\
%     \hline
%     Sofa & 27.9923  & 28.6974  & 28.5192 &  28.4587 &  \textbf{29.0628}\\
%     \hline
%   \end{tabular}
%   \label{total_results_snr}
%   \end{center}
% \end{table}

% \begin{table}[ht]
% \caption{Results: MSE}
%   \begin{center}
%   \begin{tabular}{|c|c|c|c|c|c|}
%   \hline
%     Model & BF & PDE & GRAPH & NLD & WMP\\
%     % 0.3121    0.2565    0.2754    0.2769    0.2503
%     \hline
%     Bottle & 0.2077 & 0.1572 & 0.1642 & 0.1560 & \textbf{0.1440
% } \\
%     \hline
%     Bowl & 0.3514 & 0.2839 & 0.3007 & 0.2933 & \textbf{0.2644} \\
%     \hline
%     Can & 0.3612  &  0.2898 &   0.3026  &  0.2960  & \textbf{0.2675} \\
%     \hline
%     Cap & 0.2932 &   0.2302  &  0.2419  &  0.2357  &  \textbf{0.2130} \\
%     \hline
%     Chair & 0.3121 & 0.2565 & 0.2754 & 0.2769 & \textbf{0.2503} \\
%     \hline
%     Guitar & 0.2164  &  0.1726  &  0.1762  &  0.1758  &  \textbf{0.1651} \\
%     \hline
%     Pillow & 0.2929  &  0.2349  &  0.2474  &  0.2396   & \textbf{0.2162}\\
%     \hline
%     Plane & 0.1564 &  0.1253 &  0.1315  & 0.1324  & \textbf{0.1209}\\
%     \hline
%     Sofa & 0.2904  &  0.2468  &  0.2570  &  0.2609  &  \textbf{0.2268} \\
%     \hline
%   \end{tabular}
%   \label{total_results_mse}
%   \end{center}
% \end{table}

% \begin{table}[ht]
% \caption{Results: Chamfer Distance}
%   \begin{center}
%   \begin{tabular}{|c|c|c|c|c|c|}
%   \hline
%     Model & BF & PDE & GRAPH & NLD & WMP\\
%     % 0.7153    0.4129    0.4679    0.4063    0.3706
%     \hline
%     Bottle & 0.3737 & 0.1647 & 0.1972 & 0.1440 & \textbf{0.1227
% } \\
%     \hline
%     Bowl & 0.7153 & 0.4129 & 0.4679 & 0.4063 & \textbf{0.3706} \\
%     \hline
%     Can & 0.7631  &  0.4206  &  0.4800  &  0.3974  &  \textbf{0.3639} \\
%     \hline
%     Cap & 0.5502 &   0.2802 &   0.3177 &   0.2660  &  \textbf{0.2397}\\
%     \hline
%     Chair & 0.6081 &   0.4030  &  0.4665  &  0.4539  &  \textbf{0.4011}\\
%     \hline
%     Guitar & 0.3092  &  0.1702  & 0.1841 & 0.1667  & \textbf{0.1507}\\
%     \hline
%     Pillow & 0.6085  &  0.3324 &   0.3897   & 0.3192 &   \textbf{0.2832} \\
%     \hline
%     Plane & 0.2543  & 0.1475  & 0.1647  & 0.1552  & \textbf{0.1425}\\
%     \hline
%     Sofa & 0.7514  &  0.5127  &  0.5597  &  0.5187  &  \textbf{0.4731} \\
%     \hline
%   \end{tabular}
%   \label{total_results_cd}
%   \end{center}
% \end{table}

% \begin{figure*}[!htb]
% 	\begin{minipage}[t]{0.135\linewidth}
% 		\centering
% 		\includegraphics[width=1.3in]{sofa_pure.jpg}
% 		\caption*{(a) Original PC} 
% 	\end{minipage}%
% 	\begin{minipage}[t]{0.135\linewidth}
% 		\centering
% 		\includegraphics[width=1.3in]{sofa_noise.jpg}
% 		\caption*{(b) Noisy PC} 
% 	\end{minipage}
% 	\begin{minipage}[t]{0.135\linewidth}
% 		\centering
% 		\includegraphics[width=1.3in]{sofa_bf.jpg}
% 		\caption*{(c) BF} 
% 	\end{minipage}
% 	\begin{minipage}[t]{0.135\linewidth}
% 	\centering
% 	\includegraphics[width=1.3in]{sofa_graph.jpg}
% 	\caption*{(d) GD} 
% 	\end{minipage}
%     \begin{minipage}[t]{0.135\linewidth}
% 	\centering
% 	\includegraphics[width=1.3in]{sofa_nld.jpg}
% 	\caption*{(e) NLD} 
% 	\end{minipage}
%     \begin{minipage}[t]{0.135\linewidth}
% 	\centering
% 	\includegraphics[width=1.3in]{sofa_pde.jpg}
% 	\caption*{(f) PDE} 
% 	\end{minipage}
%     \begin{minipage}[t]{0.135\linewidth}
% 	\centering
% 	\includegraphics[width=1.3in]{sofa_wmp.jpg}
% 	\caption*{(g) WMP} 
% 	\end{minipage}
% 	\caption{Sofa} \label{mesh_sofa}
% \end{figure*}

\section{Conclusions}\label{conclusion}
We proposed a novel algorithm for 3D point cloud denoising by averaging the projections from multiple tangent planes. We analyzed the error of estimated normal vector and achieve a tighter bound compared to a previous work. We also showed that the multi-projection  is more robust than the one-time projection. We validate the proposed algorithm on real datasets and compared with other 3D point cloud denoising algorithms and show that our algorithm outperforms its competitors in all nine categories for all three evaluation metrics.
\newpage

\section*{Appendix}
\subsection{Davis Kahan theorem}
Recall that if $V , \hat{V}\in R^{p\times d}$ both have orthonormal columns, then the vector of $d$ principal angels between their column spaces is give by
\begin{equation}
(\cos^{-1}(\sigma_1) , \cdots , \cos^{-1}(\sigma_d))^T,
\end{equation}
where $\sigma_1 \geq \cdots \geq \sigma_d$ are the singular values of $\hat{V}^TV$. Let $\Theta (\hat{V} , V)$ denote the $d \times d$ diagonal matrix whose $j$th diagonal entry is the $j$th principal angle, and let $\sin \Theta(\hat{V} , V)$ be defined entrywise.

\textbf{Davis Kahan theorem}: Let $\Sigma , \hat{\Sigma}\in R^{p\times p}$ be symmetric, with eigenvalus $\lambda_1 \geq \cdots \geq \lambda_p$ and $\hat{\lambda}_1 \geq \cdots \geq \hat{\lambda}_q$ respectively. Fix $1 \leq r \leq s \leq p$, let $d := s - r + 1$, and let $V = (\mathbf{v}_r , \mathbf{v}_{r+1} , \cdots , \mathbf{v}_s)\in \mathbb{R}^{p\times d}$ and $\hat{V} = (\hat{\mathbf{v}}_r , \hat{\mathbf{v}}_{r+1} , \cdots , \hat{\mathbf{v}}_s) \in \mathbb{R}^{p\times d}$ have orthonormal columns satisfying $\Sigma v_j = \lambda_j v_j$ and $\hat{\Sigma}\hat{v}_j = \hat{\lambda}_j\hat{v}_j$ for $j = r , r+1 , \cdots , s$. If $\delta := \inf \{|\hat{\lambda} - \lambda| : \lambda \in [\lambda_s , \lambda_r ], \hat{\lambda} \in (-\infty , \hat{\lambda}_{s-1}] \cup [\hat{\lambda}_{r+1} , \infty ) \} > 0$, where $\hat{\lambda}_0 := -\infty$ and $\hat{\lambda}_{p+1} := \infty$, then
\begin{equation}
	\| \sin \Theta (\hat{V} , V)\|_F \leq \frac{\| \hat{\Sigma} - \Sigma\|_{F}}{\delta}.
\end{equation}
Frequently in applications, we have $r = s = j$, in which case we can conclude that
\begin{equation}
|\sin \Theta (\hat{\mathbf{v}}_j , \mathbf{v}_j)| \leq \frac{\| \hat{\Sigma} - \Sigma\|_{op}}{\min(|\hat{\lambda}_{j-1} - \lambda_j| , |\hat{\lambda}_{j+1} - \lambda_j|)}
\end{equation}

\subsection{Covariance Matrix For Noise-free Point Cloud}
For the noise-free point cloud based on the model in Sec. \ref{math}, we have
$$
\mathbf{p}_i \ = \ \begin{bmatrix}
x_i \\ y_i
\end{bmatrix}
= \begin{bmatrix}
r\sin \theta_i \\ r - r\cos \theta_i
\end{bmatrix},
$$
and the mean values of coordinates of the sampled points are as follows,
\begin{equation}
	\begin{aligned}
	\bar{x} & = 0, \\
	\bar{y} & = r(1 - \frac{\sin \alpha}{\alpha}),\\
    \alpha & = \frac{\epsilon N}{r}.
	\end{aligned}
\end{equation}
To simplify the analysis, let 
$$y(i) = y(i) - \bar{y},\quad
y_n(i) = y_n(i) - \bar{y}.$$ 
The covariance matrix for origin point $\mathbf{p}_0$ is
\begin{equation}
	\begin{aligned}
	M_o & = \sum_{j = -N}^{N}(\left[
	\begin{matrix}
	x(j) \\
	y(j) 
	\end{matrix}
	\right] - \left[\begin{matrix}
	\bar{x}\\ \bar{y}
	\end{matrix}\right])
	(\left[
	\begin{matrix}
	x(j) \\
	y(j) 
	\end{matrix}
	\right] - \left[\begin{matrix}
	\bar{x}\\ \bar{y}
	\end{matrix}\right])^T\\
	& = \sum_{j = -N}^{N}\left[\begin{matrix}
	x(j) \\ 
	y_r(j)
	\end{matrix}\right]
	\left[\begin{matrix}
	x(j) \quad y(j)
	\end{matrix}\right] \\
	& = \left[\begin{matrix}
	m_{11} & m_{12} \\
	m_{12} & m_{22}
	\end{matrix}\right] = \left[\begin{matrix}
	m_{11} & 0 \\
	0 & m_{22}
	\end{matrix}\right].
	\end{aligned}
\end{equation}
Since we know $0< \alpha < \frac{\pi}{2}$, we know $m_{11} > m_{12}$. The eigenvalues of this matrix is $\lambda_1 = m_{11}, \lambda_2 = m_{22}$. The corresponding eigenvectors are $v_1 = [1 \quad 0], v_2 = [0 \quad 1]$. 

\subsection{Covariance Matrix For Noisy Point Cloud}
For the noisy point cloud, we have 
$$\widetilde{\mathbf{p}}_i \ = \ \begin{bmatrix}
\widetilde{x}_i \\ \widetilde{y}_i
\end{bmatrix}
= \begin{bmatrix}
r\sin \theta_i + n_{x,i}
\\ 
r - r\cos \theta_i + n_{y,i}
\end{bmatrix}.
$$
The covariance matrix for point $\widetilde{\mathbf{p}}_0$ is
\begin{equation}
	\begin{aligned}
	\widetilde{M}_o & = \sum_{j = -N}^{N}(\left[
	\begin{matrix}
	x_n(j) \\
	y_n(j) 
	\end{matrix}
	\right] - \left[\begin{matrix}
	\overline{x}\\ \overline{y}
	\end{matrix}\right])
	(\left[
	\begin{matrix}
	x_n(j) \\
	y_n(j) 
	\end{matrix}
	\right] - \left[\begin{matrix}
	\bar{x}\\ \bar{y}
	\end{matrix}\right])^T\\
	& = \sum_{j = -N}^{N}\left[\begin{matrix}
	x_n(j) \\ 
	y_n(j)
	\end{matrix}\right]
	\left[\begin{matrix}
	x_n(j) \quad y_n(j)
	\end{matrix}\right] \\
	& = \left[\begin{matrix}
	\widetilde{m}_{11} & \widetilde{m}_{12} \\
	\widetilde{m}_{12} & \widetilde{m}_{22}
	\end{matrix}\right].
	\end{aligned}
\end{equation}

\subsection{Error Estimation}
Based on Davis Kahan Theorem, we still have
\begin{equation}
	|\sin \Theta (\hat{\mathbf{v}}_j , \mathbf{v}_j)| \leq \frac{\| \hat{\Sigma} - \Sigma\|_{op}}{m_{11}},
\end{equation}
where $\hat{\Sigma} = \hat{M}_c$ and $\Sigma = M_c$. We know
\begin{equation}
\begin{aligned}
	\hat{\Sigma} - \Sigma  & = \left[\begin{matrix}
	0 & \hat{m}_{12} \\
	\widetilde{m}_{12} & \widetilde{m}_{22} - m_{22}
	\end{matrix}\right], \\
	\therefore \quad |\sin(\theta)| & \leq \frac{|\widetilde{m}_{12}| + |\widetilde{m}_{22} - m_{22}| }{m_{11}}
\end{aligned}
\end{equation}

\newpage

%\section*{References}
\bibliographystyle{IEEEtran}
\bibliography{0my_ref}

\end{document}